\documentclass{JAC2003}

\usepackage{graphicx}

\begin{document}
\title{TESTS OF A ROMAN POT PROTOTYPE FOR THE TOTEM EXPERIMENT\vspace*{-5mm}}
\author{M. Deile, E. Alagoz, G. Anelli, G. Antchev, M. Ayache, 
        F. Caspers, E. Dimovasili, R. Dinapoli,\\ 
F. Drouhin, K. Eggert, J.L. Escourrou, O. Fochler, K. Gill, R. Grabit, 
F. Haug, P. Jarron, J. Kaplon,\\ 
T. Kroyer, T. Luntama, D. Macina, E. Mattelon, H. Niewiadomski, L. Mirabito,
E.P. Noschis,\\ 
M. Oriunno, A. Park, A.-L. Perrot, O. Pirotte, J.M. Quetsch, F. Regnier, 
G. Ruggiero, S. Saramad,\\ 
P. Siegrist, W. Snoeys, T. Souissi, R. Szczygiel,
J. Troska, F. Vasey, A. Verdier (CERN, Geneva), \\
C. Da Vi\`{a}, J. Hasi, A. Kok, S. Watts (Brunel University, Middlesex),
J. Ka\v{s}par, V. Kundr\'{a}t, \\
M.V. Lokaj\'{\i}\v{c}ek,
J. Smotlacha (FZU, Prague), V. Avati, M. J\"{a}rvinen, 
M. Kalliokoski, J. Kalliopuska, \\
K. Kurvinen,
R. Lauhakangas, F. Oljemark, R. Orava, K. \"{O}sterberg,
V. Palmieri, H. Saarikko,\\
A. Soininen (Helsinki University), V. Boccone, M. Bozzo, A. Buzzo,
S. Cuneo, F. Ferro, M. Macr\'{\i}, \\
S. Minutoli, A. Morelli, 
P. Musico, M. Negri, A. Santroni, G. Sette,
A. Sobol (INFN Genova), \\
V. Berardi, M.G. Catanesi, 
E. Radicioni (INFN-Bari)}
\maketitle

\begin{abstract}
The TOTEM collaboration has developed and tested the first prototype of its
Roman Pots to be operated in the LHC. TOTEM Roman Pots contain
stacks of silicon detectors with strips oriented in two orthogonal
directions. To measure proton scattering angles of a few microradians, the 
detectors will approach the beam centre to a distance of 
$10\,\sigma + 0.5$\,mm (= 1.3 mm). Dead space near the detector edge is
minimised by using two novel ``edgeless'' detector technologies.
The silicon detectors are used both for precise track reconstruction and 
for triggering. The first full-sized prototypes of both detector technologies 
as well as their read-out electronics have been developed, built and operated.
The tests took place in the proton beam-line of the SPS accelerator ring.
In addition, the pot's shielding against electromagnetic interference 
and the longitudinal beam 
coupling impedance have been measured with the wire method.
\end{abstract}

\section{Roman Pots for the TOTEM Experiment at LHC}
The LHC experiment TOTEM~\cite{tdr} is designed for measuring the elastic pp scattering
cross-section, the total pp cross-section and diffractive processes. These
physics objectives require the detection of leading protons with scattering
angles of a few $\mu$rad, which is accomplished with a Roman Pot (``RP'') system having
stations at 147\,m and 220\,m 
from the interaction point 5 where CMS will be located.
Each station is composed of two RP units separated by 2.5 -- 4\,m
depending on beam equipment integration constraints.
Each RP unit consists of a vacuum chamber equipped with two 
vertical insertions (top and bottom) and a horizontal one. 
Each insertion (``pot'') contains a package of 10 silicon detectors in a 
secondary vacuum. The pots can be moved into the 
primary vacuum of the machine through vacuum bellows. 
In order to minimize the distance of the detectors from the beam, and to minimize multiple scattering, the wall 
thickness of the pot is locally reduced to a thin window foil 
(140 -- 210\,$\mu$m).

The low impedance budget of the LHC machine (broadband
longitudinal~impedance limit $Z/n \approx 0.1\,\Omega$) imposes a tight  
limit on the RPs' beam coupling impedance. 
Because of the beam's bunch structure and its high
intensity, the pick-up noise on the detector electronics caused by 
electromagnetic leakage is a potential concern and needs to be studied and
minimised.

\section{The Roman Pot Prototype in the SPS Beam}
During the summer 2004 the TOTEM collaboration has performed 
tests on a prototype unit made of 2 vertical RPs. The unit
has been installed in the LSS5 section of the SPS accelerator at CERN. 
The motors moving the pots towards the beam, the electronics of the detectors 
and the cooling system stabilising the temperature of the detectors inside 
the pots were operated remotely from a temporary control room in a surface 
building. 
The upper and lower pots contained each a set of 8 silicon strip
detectors with 66\,$\mu$m pitch, built in two novel technologies allowing for 
efficiency up to the physical edge~\cite{tdr}. Six of them were read out with analogue APV25 chips 
and 2 with digital VFAT chips delivering the fast-or signal of all 512 strips.
The latter were
used for triggering the data acquisition system in coincidence with the 
sum signal of the four pick-up electrodes of a beam position monitor.

In the retracted pot position the thin windows were       
40\,mm away from the beam pipe axis.
To watch the effect of the RPs' movements on the beam quality,
three beam loss monitors (BLMs) have been installed near the unit: one at
56\,cm upstream on the top of the beam pipe and two at 65\,cm
downstream, one on the top and one on the bottom of the pipe. 
The BLMs used were cylindrical ionisation chambers filled with air. 

TOTEM was the SPS main user during 2 shifts of 8 hours with dedicated coasting 
beam. Three bunch structures were tested: 1 single bunch in the ring, 
4 bunches equally spaced, and 4 equally spaced trains of 4 bunches. 

Detector data were taken with the two pots positioned independently 
between 6\,mm and 14\,mm 
($\sigma_{beam} \approx 0.8$\,mm) from the beam pipe centre. Because of contact 
problems with the Kapton connections between the individual detector/chip 
hybrids and the central electronics board of each pot, 4 out of the 16
detectors could not be read out. The high redundancy in the system's
design however allowed us to carry out the experiment as planned.

\begin{figure}[h!]
\centering
\vspace*{-3mm}
\includegraphics*[width=70mm]{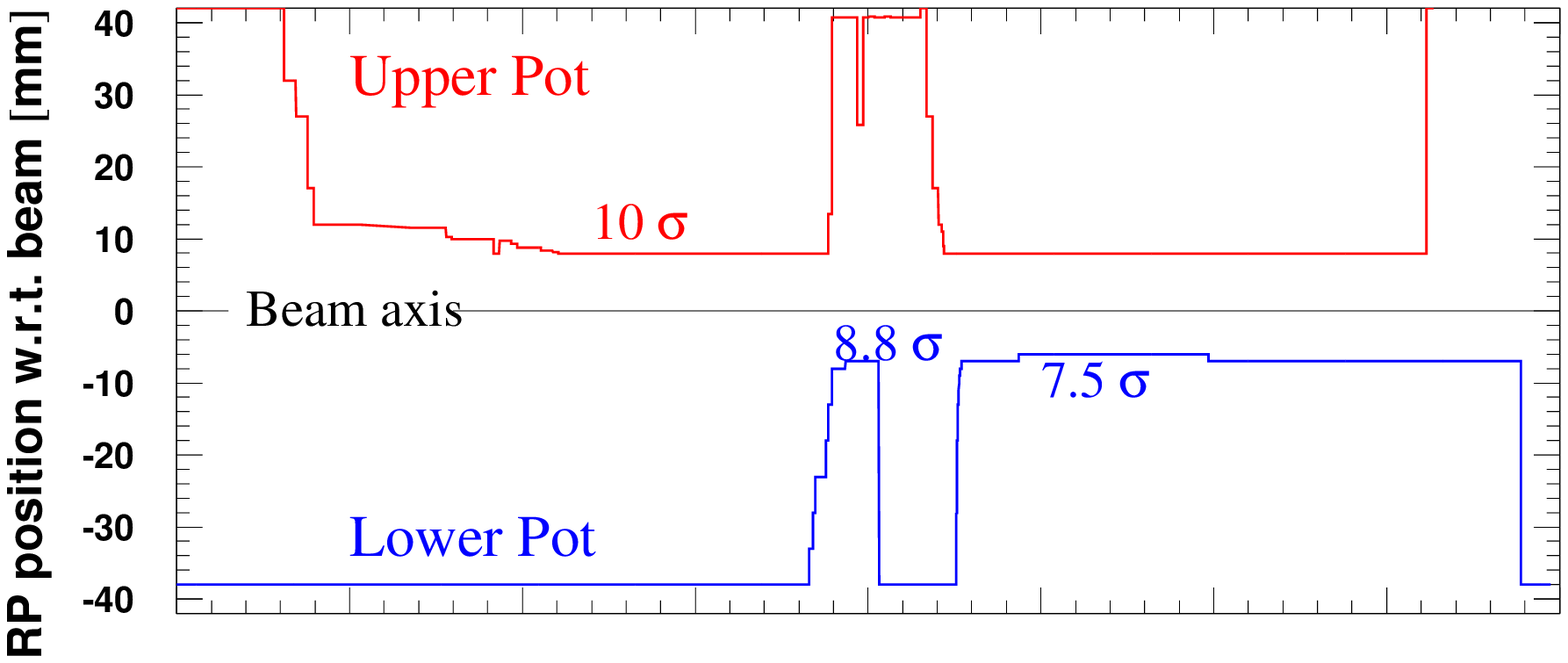}
\includegraphics*[width=70mm]{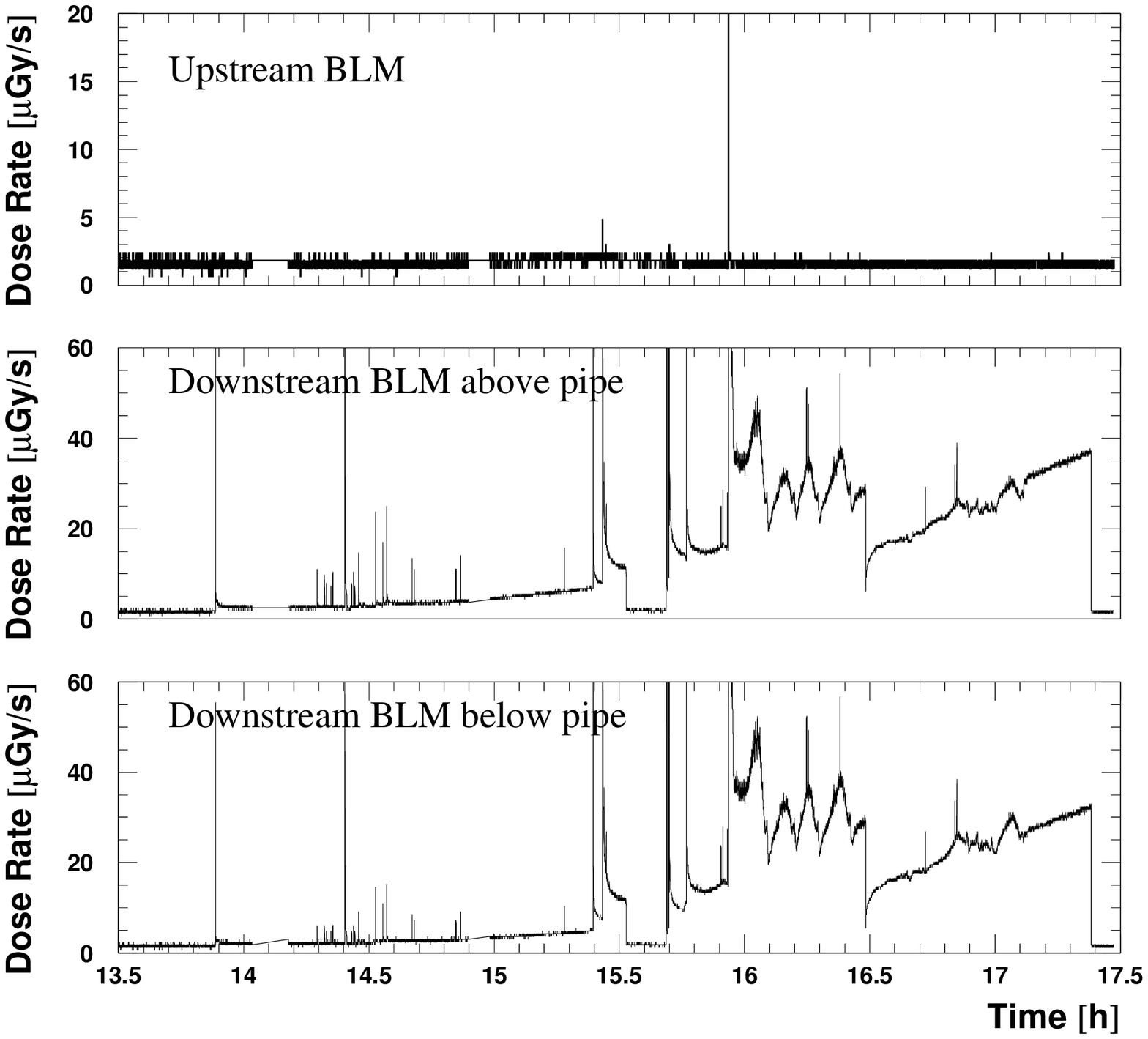}
\vspace*{-4mm}
\caption{Synchronous measurements of the RP positions 
and the dose rates in the beam loss 
monitors over the duration of a data taking period.}
\label{fig_blm}
\vspace*{-2mm}
\end{figure}
Figure~\ref{fig_blm} shows the scraping of the beam periphery by the pot
as measured with the BLMs. The downstream monitors show spikes whenever one
of the pots approaches the pipe centre closer than 10\,mm ($ = 12.5\,\sigma$). 
As expected, the 
upstream BLM remains largely quiet, except when the bottom pot moves closer 
than $8\,\sigma$, which creates an increased halo travelling around the ring. 
The general trend of increasing beam losses is caused by a slow beam growth
as its quality deteriorates.

\begin{figure}[h!]
\centering
\vspace*{-1mm}
\includegraphics*[width=70mm]{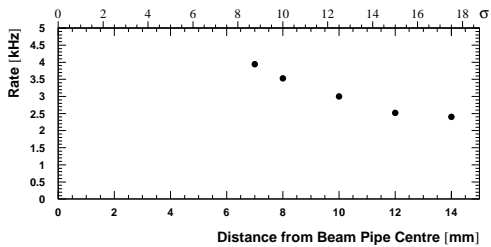}
\vspace*{-4mm}
\caption{Trigger rate as a function of the window's distance from the beam 
pipe centre.}
\label{fig_trigger}
\vspace*{-2mm}
\end{figure}
The trigger rates measured in the top pot for different distances are shown in
Figure~\ref{fig_trigger}. Given the SPS revolution period of 23\,$\mu$s, the 
trigger probability
varies between one trigger per 11 bunches at $d = 7\,$mm and one trigger per 18
bunches at $d = 14\,$mm. The 
latter distance corresponds to 17\,$\sigma$, as also foreseen for the LHC at 
the 220\,m station. The observed variation of the rate with the distance is
not big because all measurements were made more than 8\,$\sigma$ from the
beam centre where the halo is rather flat.

\begin{figure}[h!]
\vspace*{-3mm}
\centering
\hspace*{-10mm}\includegraphics*[width=95mm]{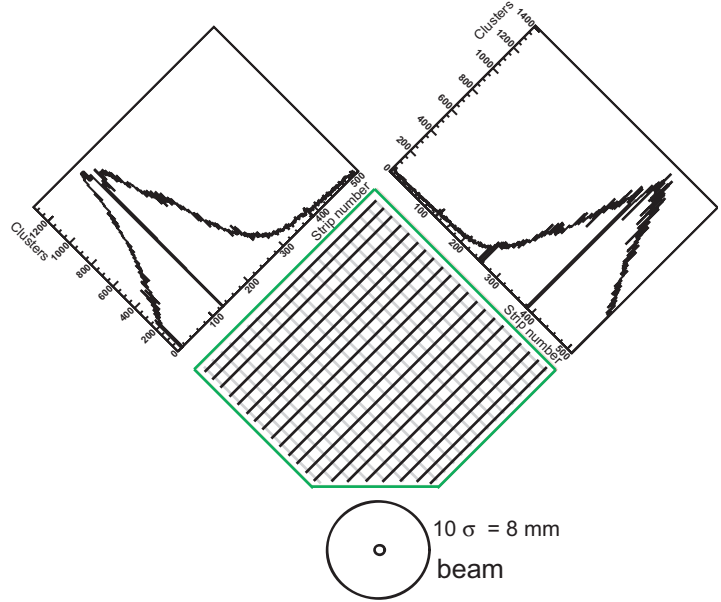}
\vspace*{-7mm}
\caption{Profile of the beam halo as seen by two detector planes
with orthogonal strip orientations at a distance of 10\,mm from the beam centre.}
\label{fig_beamprofile}
\vspace*{-2mm}
\end{figure}

The response of the tracking silicon detectors conforms to expectations.
Figure~\ref{fig_beamprofile} shows the halo profiles measured by detector 
planes with orthogonal strips.

\section{Laboratory Measurements of the RF Behaviour}
For the measurements of the pot's radiofrequency behaviour a 0.3\,mm thick 
wire was strung through the RP along its beam axis. 
\begin{figure}[h!]
\centering
\vspace*{-3mm}
\includegraphics*[width=75mm]{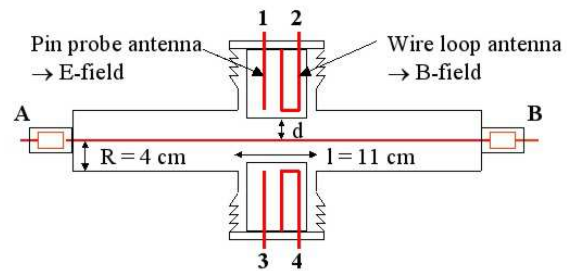}
\vspace*{-3mm}
\caption{Setup for the laboratory measurements. For the connections see text.}
\label{fig_setup}
\vspace*{-2mm}
\end{figure}

\subsection{Attenuation of Electric and Magnetic Fields by the Roman Pot Window}
\begin{figure}[h!]
\centering
\includegraphics*[width=70mm]{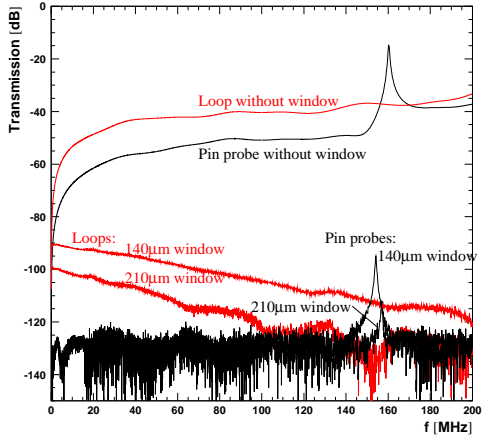}
\vspace*{-3mm}
\caption{Signal transmission from the wire to the 
antennas as a function of the frequency for a distance of 2\,mm between each pot and the wire in the centre. The peaks at 160\,MHz are caused by a resonance
of the pin probe. Corrections for reflections at Port A and for the 30\,dB 
amplification have been applied.}
\label{fig_attenvsf}
\vspace*{-4mm}
\end{figure}
\begin{figure}[h!]
\centering
\vspace*{-3mm}
\includegraphics*[width=70mm]{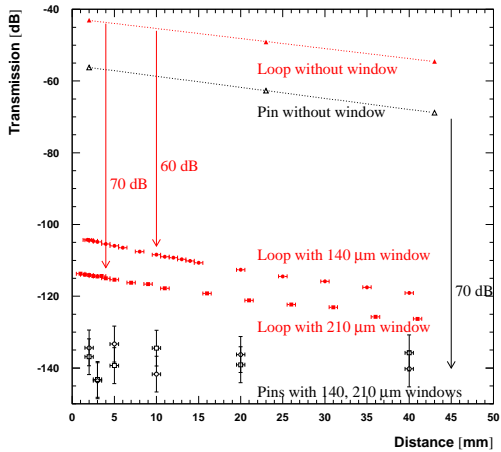}
\vspace*{-3mm}
\caption{Signal transmission from the wire to the
antennas at 40\,MHz as a function of the distance from the wire.}
\label{fig_attenvsd}
\end{figure}

To determine the attenuation of electromagnetic fields by the RP's windows, the silicon
detectors were replaced with antenna insertions (Figure~\ref{fig_setup}). 
A network analyser measured the transmission from one end of the
RP (Port A) to each of the antenna outputs 
(Ports 1 to 4) for frequencies up to 200\,MHz. 
The other end of the RP (Port B) was terminated with 50\,$\Omega$. The 
transmitted signal was amplified by 30\,dB. The top and bottom insertions had 
210\,$\mu$m and 140\,$\mu$m thick windows respectively. To measure the pure
signal coupling between the wire and the antennas, comparison measurements 
without any window have been made. The results for a distance of 2\,mm are 
shown in Figure~\ref{fig_attenvsf}. The transmitted electric field is mostly 
smaller than the noise level of the measurement setup.
Figure~\ref{fig_attenvsd} shows the transmission from the wire to the antennas
at the LHC bunch frequency 40\,MHz as a function of the pot distance from the
wire. The attenuation of the magnetic field by the 140\,$\mu$m and 210\,$\mu$m
thick pot windows amounts to 60\,dB and 70\,dB respectively. The electric field
is attenuated by at least 70\,dB. The influence of bunch-generated 
fields on detectors and electronics remains to be measured in the future with
a spark generator.

\subsection{The Longitudinal Beam Coupling Impedance}

\begin{figure}[h!]
\centering
\vspace*{-3mm}
\includegraphics*[width=70mm]{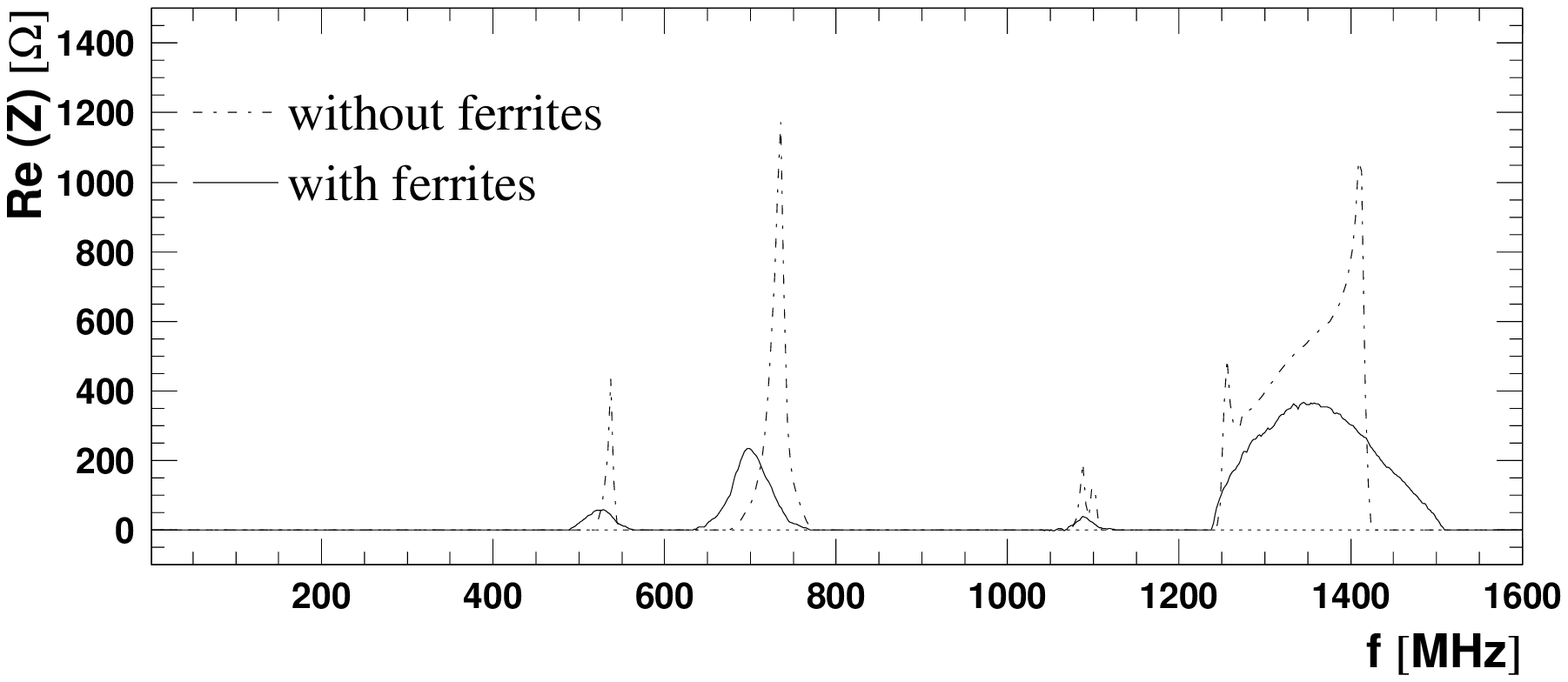}
\vspace*{-5mm}
\caption{Longitudinal beam coupling impedance for $d = 2\,$mm 
before and after installing ferrites in the cavity around the pot wall.}
\label{fig_impedance}
\vspace*{-2mm}
\end{figure}
\begin{figure}[h!]
\centering
\vspace*{-1mm}
\includegraphics*[width=70mm]{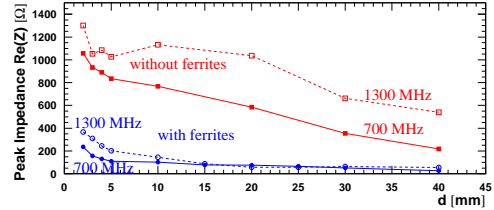}
\vspace*{-5mm}
\caption{Impedance as a function of $d$ at the
two dominant resonances.}
\label{fig_impvsdist}
\vspace*{-2mm}
\end{figure}
To determine the longitudinal beam coupling impedance $Z$, the complex
transmission coefficient $S_{21}$ between Ports A and B 
(Figure~\ref{fig_setup}) was measured with a network ana\-lyser.
Then $Z$ was calculated from $S_{21}(f,d)$ with 
the RP at a distance $d$, and a reference measurement 
$S^{ref}_{21}(f)$ with the RP in retracted position. 
The ``improved log formula'' proposed by Shaposhnikova and 
Jensen~\cite{jensen} 
\begin{equation}
Z(f,d) = -2 Z_{C}\, \ln \frac{S_{21}(f,d)}{S^{ref}_{21}(f)} \left[ 1 + i\, \frac{\ln \frac{S_{21}(f,d)}{S^{ref}_{21}(f)}}{4 \pi l\,f / c} \right]
\end{equation}
was used, where $Z_{C} = 294\,\Omega$ is the characteristic impedance of the
unperturbed beam pipe, $l = 11\,$cm is the length of the perturbation (i.e. the
diameter of the pot insertions), and $f$ is the frequency. 
The dash-dotted line in Figure~\ref{fig_impedance} shows the real part of $Z$ 
for $d = 2\,$mm with the original configuration of the RP. The dependence on
$d$ is given in Figure~\ref{fig_impvsdist} for the two main resonances.
The approximately Gaussian LHC bunch 
structure with $\sigma_t = 0.25\,$ns leads to a Gaussian envelope with 
$\sigma_f = 0.63\,$GHz in the 
frequency distribution of the LHC current with harmonics every 40\,MHz.
Hence the relevant resonances lie well below 1\,GHz. The dominant line at 
740\,MHz has an impedance of 1.2\,k$\Omega$ which yields an uncomfortably high
$Z/n = 18\,{\rm m}\Omega$ where $n = f_{reson}/f_{LHC} = 740\,{\rm MHz} / 11\,{\rm kHz}$.
After installation of little pieces of ferrite into the cavity around the 
RP insertion, this impedance is damped by a factor 5, demonstrating the 
effectiveness of the ferrite approach. For the final RP the geometrical
ferrite arrangement will be optimised to further reduce the impedance.

\end{document}